\documentclass[a4paper,aps,prd,twocolumn,preprintnumbers,showpacs,superscriptaddress,nofootinbib]{revtex4}
\usepackage{graphics,graphicx,epsfig}
\usepackage{amsfonts,amsmath,amssymb}
\usepackage{color}

\begin{document}

\title{Baby de Sitter Black Holes and dS$_3$/CFT$_2$}
\author{Sophie de Buyl}
\email{sdebuyl-at-physics.harvard.edu}
\affiliation{Center for the Fundamental Laws of Nature, Harvard University,\\
Cambridge, MA 02138, USA}
\author{St\'ephane Detournay}
\email{sdetourn-at-physics.harvard.edu}
\affiliation{Center for the Fundamental Laws of Nature, Harvard University,\\
Cambridge, MA 02138, USA}
\author{Gaston Giribet}
\email{gaston-at-df.uba.ar}
\affiliation{IFIBA-CONICET and Physics Department, University of Buenos Aires,\\
Buenos Aires, 1428 Argentina}
\author{Gim Seng Ng}
\email{gng-at-fas.harvard.edu}
\affiliation{Center for the Fundamental Laws of Nature, Harvard University,\\
Cambridge, MA 02138, USA}

\begin{abstract}
Unlike three-dimensional Einstein gravity, three-dimensional massive gravity
admits asymptotically de Sitter space (dS) black hole solutions. These black
holes present interesting features and provide us with toy models to study the
dS/CFT correspondence. A remarkable property of these black holes is that
they are always in thermal equilibrium with the cosmological horizon of the
space that hosts them. This invites us to study the thermodynamics of these
solutions within the context of dS/CFT. We study the asymptotic symmetry
group of the theory and find that it indeed coincides with the local
two-dimensional conformal algebra. The charge algebra associated to
the asymptotic Killing vectors consists of two copies of the Virasoro
algebra with non-vanishing central extension. We compute the mass and
angular momentum of the dS black holes and verify that a naive application of
Cardy's formula exactly reproduces the entropy of both the black hole and the
cosmological horizon. By adapting the holographic renormalization techniques
to the case of dS space, we define the boundary stress tensor of the dual
Euclidean conformal field theory.
\end{abstract}

\maketitle

\section{Introduction}

Three-dimensional gravity has shown to be a fruitful playground to study the
AdS/CFT holographic correspondence and its ramifications. In particular, its
features point at the universality of statistical black hole entropy
derivations when an AdS$_3$ factor is involved \cite{Strominger}. It also
appears as a promising path to explore non-AdS holography, such as
anisotropic scale-invariant condensed-matter models \cite{Waves, Lifshitz},
Flat Space holography \cite{glenn, flatCG, glenn2, detournay} and Warped AdS$
_3$ spaces \cite{Warped,1210.0539}. 
Three-dimensional gravity has also played a special role in relation to the
dS/CFT correspondence \cite{dSCFT,Hull:1998vg,Witten:2001kn, Klemm:2001ea,
Cacciatori:2001un, Maldacena:2002vr}.
It is the aim of this paper to further explore this direction.

The dS/CFT correspondence \cite{dSCFT,Hull:1998vg,Witten:2001kn,
Klemm:2001ea, Cacciatori:2001un, Maldacena:2002vr} was proposed with the aim
of extending the idea of holography to asymptotically de Sitter space. This
is certainly a problem of principal importance since we might actually live
in such a universe! The main idea is that quantum gravity on $d$-dimensional
dS space could be in correspondence with a $(d-1)$-dimensional Euclidean
conformal field theory. More precisely, mimicking what happens in AdS/CFT, a
minimal version of the dS/CFT\ conjecture states that perturbative
correlation functions in dS$_{d}$ with appropriate future boundary
conditions \cite{Anninos:2011jp,Ng:2012xp} on future null infinity $\mathcal{%
I}^{+}$ are given by correlation functions of a $(d-1)$-dimensional
Euclidean conformal field theory on $\mathcal{I}^{+}$.

In general, extracting precise information from holography is a non-trivial
task. This is even more so in dS/CFT partly due to the fact that one still does
not have a good definition of relevant observables in dS space\footnote{%
Nevertheless, see Ref.~\cite{Marolf:2012kh,Anninos:2011af,Anninos:2012qw}
for some recent discussions on this issue.}. There are, however, some
examples in which the holographic dictionary can be established in a more
precise way: For instance, a microscopic realization of dS$_{4}$/CFT$_{3}$
was proposed \cite{Anninos:2011ui} in the context of higher-spin gravity on
dS$_{4}$, relating Vasiliev's theory in dS$_{4}$ to the dual Sp(N) CFT$_{3}$%
. This has enabled further investigations\cite%
{Anninos:2012ft,Ng:2012xp,Anninos:2013rza,Banerjee:2013mca} into sharper
statements about dS$_{4}$/CFT$_{3}$ which hopefully will lead to a better
understanding of de Sitter space. For
general dS$_{d}$/CFT$_{d-1}$, on the other hand, one still hopes that
symmetry arguments would already provide some interesting physical information. This is
particularly realized in the $d=3$ case\footnote{%
Generalization of the higher-spin dS/CFT to $d=3$ case is much less
understood, although it has recently been investigated \cite{Ouyang:2011fs}.}%
, where the conformal group is infinite-dimensional. A particular
interesting application of dS$_{3}$/CFT$_{2}$ yields a highly-suggestive numerological
derivation of the Kerr-dS$_{3}$ cosmological horizon in terms of the Cardy
entropy formula for the dual CFT$_{2}$ \cite{Park:1998qk, Bousso:2001mw}.

Besides pure Einstein gravity (GR), other gravitational theories in 2+1 dimensions (often involving higher-derivative terms) have been explored and have attracted a lot of attention in recent years.
An interesting class of such examples includes models
of massive gravity. These theories generally contain local propagating
degrees of freedom, but this has a price: Higher-curvature models usually
lead to inconsistencies already at the perturbative level due to the
presence of negative-norm (or energy) gravitons. To fix this, one needs to
decouple such states coming from extra degrees of freedom due to the higher
derivatives. A particularly nice class of theories where such decoupling can
be accomplished is the Bergshoeff-Hohm-Townsend (BHT) model\cite{NMG,NMG2}
(also known as New Massive Gravity or NMG), which amounts to supplementing the
Einstein-Hilbert action with special $R^{2}$-terms. Linearized gravitational
perturbations in BHT model around flat, AdS$_{3}$ or dS$_{3}$ space have
been shown to be free of ghost/negative norm states\cite{NMG2} at some
special values of the couplings.

The general motivation to study the BHT theory of massive gravity is that it
provides a toy-model for four-dimensional Einstein gravity. In fact, BHT
exhibits several properties that are reminiscent of four-dimensional GR: For
instance, in both of these theories there exist two well-behaved local
propagating degrees of freedom. Furthermore, in BHT theory there exist
interesting non-trivial black holes both in AdS$_{3}$ and in dS$_{3}$ spaces
\cite{GOTT,OTT,Gabadadze:2012xv}. These black holes generally have finite
mass and angular momentum, together with an extra hair parameter. The causal
structure of the (A)dS$_{3}$ black holes of BHT theory resembles that of the
(A)dS$_{4}$ black holes of GR: the Penrose diagrams are similar, and in
these geometries there exists a curvature singularity at the origin,
shielded by the horizon. Therefore, it provides an interesting playground to
apply holographic techniques and try to understand how to interpret hairy
black holes with curvature singularities in the dual theory. Some attempts
have been made along these lines by studying the scalar wave equation in these
backgrounds \cite{Kwon:2011ey}, which can be written as the Heun
differential equation. This type of differential equation also appears in
studying perturbations about four-dimensional black holes.

As mentioned, in this paper we aim to extend the analysis of dS$_{3}$/CFT$%
_{2}$ correspondence to the BHT theory and, at the same time, consider black
holes in dS$_{3}$. This analysis will then resemble what might happen in
higher-dimensional dS$_{d}$/CFT$_{d-1}$ where black holes do exist. We will
study the asymptotic symmetry group of de Sitter space in this theory and
show that, indeed, it is generated by two copies of the Virasoro algebra
with non-trivial central extensions. This work differs from the dS$_{3}$
asymptotic symmetries analysis of \cite{dSCFT} in that we are no longer
working in the context of GR. As a consequence, we are considering a more
relaxed set of boundary conditions to include the hairy black hole
solutions. This results in a different central charge. Despite the more
relaxed asymptotic behavior, we will find that the black hole solutions have
finite conserved charges which are consistent with the first law of black hole mechanics. One curious
feature of these black holes (which already occurred in the AdS$_{3}$ case%
\cite{OTT,GOTT}) is that they only possess two horizons and both horizons
share \textit{exactly} the same thermodynamical properties. In dS space this
implies that the black holes are always in thermal equilibrium with respect
to the cosmological horizon. We will find that a naive application of the
Cardy formula exactly reproduces the entropy of \textit{both} the dS
cosmological horizon of the black hole horizon.

The paper is organized as follows: In Section II we discuss
three-dimensional massive gravity in asymptotically dS spaces. We analyze the
asymptotic isometry group and show it is generated by two copies of Witt
algebra, extending the sl(2,$\mathbb{C}$) isometry algebra. We also show
that the charges associated to the asymptotic Killing vectors close an
algebra that turns out to be two copies of Virasoro algebra with negative
central extension. In Section III we consider asymptotically dS$_{3}$ black
hole solutions of BHT theory. These black holes appear at the point of the
parameter space at which the BHT theory becomes partially massless. For
these solutions we study the conserved charges using both the canonical
approach and the quasilocal stress-tensor defined at the boundary of the
spacetime. In Section IV we study the thermodynamics both of the dS$_{3}$
black holes and of the cosmological horizon. We show that the first law of
black hole mechanics holds in BHT theory on dS$_{3}$. Then, we study the
extremal limit of these black holes as well as their local thermodynamic
(in)stability. We also show that the dS$_{3}$ black hole thermodynamics can
be reproduced from the dual CFT$_{2}$ point of view. More precisely, a
direct application of the Cardy formula is shown to exactly match both the
black hole and the cosmological horizon entropy formulas. This is remarkable
since, in addition to mass and angular momentum, the dS$_{3}$ black holes
have an extra hair parameter which eventually combines with the charges in
such a way that matching holds. Section V contains our conclusions.

\section{Massive gravity in de Sitter space}

\subsection{Three-dimensional Massive gravity}

Consider the three-dimensional massive gravity theory defined by the action%
\footnote{%
Our convention for the epsilon tensor is $\epsilon _{\phi t r}=\sqrt{-g}$.}
\begin{eqnarray}
I &=&\frac{1}{16\pi G}\int_{\Sigma } d^{3}x\sqrt{-g}\left( R-2\Lambda
\right) +  \notag \\
&&+\frac{1}{32\pi G\mu }\int_{\Sigma } d^{3}x\sqrt{-g}\epsilon ^{\mu \nu
\sigma }\Gamma _{\mu \beta }^{\eta }\left( \partial _{\nu }\Gamma _{\eta
\sigma }^{\beta }+\Gamma _{\nu \rho }^{\beta }\Gamma _{\eta \sigma }^{\rho
}\right) +  \notag \\
&&+\frac{1}{16\pi Gm^{2}}\int_{\Sigma } d^{3}x\sqrt{-g}\left( R_{\mu \nu
}R^{\mu \nu }-\frac{3}{8}R^{2}\right) .  \label{SW}
\end{eqnarray}

The first line in (\ref{SW}) corresponds to the usual Einstein-Hilbert
action, while the second line corresponds to the gravitational Chern-Simons
term of Topologically Massive Gravity (TMG) \cite{TMG}; the third line
includes the $R^{2}$-terms of the New Massive Gravity (NMG) proposed in Ref.
\cite{NMG}, hereafter referred to as BHT theory.

Action (\ref{SW}) defines a theory of massive gravity in $d=3$ dimensions,
which is (perturbatively) ghost-free about its maximally symmetric vacua for
some values of the couplings and with appropriate choice of boundary
conditions. The field equations coming from (\ref{SW})\ read
\begin{equation}
E_{\mu \nu} := R_{\mu \nu }-\frac{1}{2}Rg_{\mu \nu }+\Lambda g_{\mu \nu }+%
\frac{1}{\mu }C_{\mu \nu }+\frac{1}{2m^{2}}K_{\mu \nu }=0,  \label{fieldeq}
\end{equation}%
where $C_{\mu \nu }$ is the Cotton tensor%
\begin{equation}
C_{\mu \nu }=\epsilon _{\mu }^{\ \alpha \beta }\nabla _{\alpha }\left(
R_{\beta \nu }-\frac{1}{4}g_{\beta \nu }R\right) ,  \label{Ct}
\end{equation}%
and where the tensor $K_{\mu \nu }$ is given by
\begin{eqnarray}
K_{\mu \nu } &=&2\square {R}_{\mu \nu }-\frac{1}{2}\nabla _{\mu }\nabla
_{\nu }{R}-\frac{1}{2}g_{\mu \nu }\square {R}+4R_{\mu \alpha \nu \beta
}R^{\alpha \beta }-  \notag \\
&&\frac{3}{2}RR_{\mu \nu }-R_{\alpha \beta }R^{\alpha \beta }g_{\mu \nu }+%
\frac{3}{8}R^{2}g_{\mu \nu }.  \label{Kt}
\end{eqnarray}

The property that makes the theory defined by action (\ref{SW}) of interest
is the absence of the ghostly excitation associated to the mode $\square {R}$%
. This follows from the fact that tensors (\ref{Ct}) and (\ref{Kt}) satisfy
the relations%
\begin{equation*}
C_{\mu }^{\ \mu }=0,\quad \quad K_{\mu }^{\ \mu }=R_{\mu \nu }R^{\mu \nu }-%
\frac{3}{8}R^{2},
\end{equation*}%
which makes the dependence in $\square {R}$ to disappear from the trace of the
equations of motion (\ref{fieldeq}).

Here, we restrict ourselves to the parity preserving model, i.e. $1/\mu =0$.
Nevertheless, it is worth mentioning that all the solutions we will discuss
here persist when the gravitational Chern-Simons term is included \cite{OTT2}%
.

Equations (\ref{fieldeq}) admit asymptotically (anti-) de Sitter solutions
with effective (A)dS length $l=l_{\pm }$ where%
\begin{equation}
l_{\pm }^{2}\equiv \frac{1}{2\Lambda }\left( 1\pm \sqrt{1+\Lambda m^{-2}}%
\right) .  \label{GGJ}
\end{equation}

Here, we will be concerned with the case of positive $l$. In that case, the
maximally symmetric solution is given by%
\begin{equation}
ds^{2}=-\left( -\frac{r^{2}}{l^{2}}+1\right) dt^{2}+\left( -\frac{r^{2}}{%
l^{2}}+1\right) ^{-1}dr^{2}+r^{2}d\phi ^{2},  \label{EldS}
\end{equation}%
which corresponds to dS$_{3}$ space. The six Killing vectors are
\begin{eqnarray}  \label{b1}
J_{0}^{\pm } &=&-\frac{1}{2}(l\partial _{t}\pm i\partial _{\phi }),\quad
\label{a1} \\
J_{\sigma }^{\pm } &=&\frac{1}{2}e^{\frac{\sigma }{l}(t\mp il\phi )}\sqrt{%
r^{2}-l^{2}}\left( \sigma \partial _{r}-\frac{rl}{r^{2}-l^{2}}\partial
_{t}\mp i\frac{1}{r}\partial _{\phi }\right)  \notag \\
\end{eqnarray}%
for $\sigma =\pm 1$. These vectors generate the sl(2,$\mathbb{C}$) algebra
\begin{equation}
\lbrack J_{1}^{+},J_{-1}^{+}]=2J_{0}^{+},\quad \lbrack J_{0}^{+},J_{\pm
1}^{+}]=\mp J_{\pm 1}^{+},
\end{equation}%
with $(J_{n}^{+})^{\ast }=J_{n}^{-}$ and hence the commutators for $%
J_{n}^{-} $'s are the same as above but with minus superscripts, while the
commutators involving one plus and one minus generator are always zero.

In this paper we will be concerned with the following special point in
parameter space:
\begin{equation}
m^{2}l^{2}=-\frac{1}{2},  \label{dD}
\end{equation}%
at which the theory exhibits special features: At (\ref{dD}), which happens
when $l_{+}^{2}=l_{-}^{2}$ in (\ref{GGJ}) and $\Lambda =-m^{2},$ the BHT
theory on dS space becomes partially massless \cite{DeNe, Hi}. In fact, this
corresponds to the so-called Deser-Nepomechie point, where massive spin-2
field theory on dS exhibits a symmetry enhancement and its local degrees of
freedom are reduced by one \cite{DeWa}. This symmetry enhancement for BHT
theory at (\ref{dD}) has been studied in \cite{NMG2, BlCv} and its relation
to the existence of dS$_{3}$ black hole has been investigated in \cite%
{Gabadadze:2012xv}. It might be of interest to explore the implications or
interpretations of this point in the context of dS/CFT in the future. Here,
we will focus on the black holes solutions which exist at (and only at) the
special point (\ref{dD}). 

Notice that since $\Lambda =1/(2l^{2})>0$, imposing $\Lambda =-m^{2}$
requires a negative value of the coupling constant $1/m^{2}$ of the $R^{2}$%
-terms in (\ref{SW}). To this regard, it is convenient to compare our
notation with that of Ref. \cite{NMG2} where this special point is also
considered. Our effective cosmological constant $l^{-2}$ is the $\Lambda $
used in \cite{NMG2}, while our cosmological parameter $\Lambda $ in the
action coincides with the product of parameters $\lambda m^{2}$ in \cite%
{NMG2}. In \cite{NMG2} a parameter $\sigma =\pm 1$ is introduced multiplying
the scalar curvature invariant in the Einstein-Hilbert piece of the action,
while in our case we have $\sigma =+1$; nevertheless, we can effectively
change from $\sigma =+1$ to $\sigma =-1$ by changing the sign of the Newton
constant $G\rightarrow -G$ and simultaneously changing $m^{2}\rightarrow
-m^{2}$. In the notation of \cite{NMG2} the partially massless point
corresponds to $\lambda =-1$, while for us this means $m^{2}l^{2}=-1/2<0$.
Notice that the choice $m^{2}<0$ with $\sigma =+1$ is actually consistent
with the dS vacuum, cf. Eq. (1.11) of \cite{NMG2}. It would be interesting
to perform the linear stability analysis about this background, which is not
included in that previous analysis.

\subsection{Phase Space, Symmetries and Charges}

The symmetries of asymptotically dS$_{3}$ spaces are generated by two copies
of Witt algebra \cite{dSCFT, Marolf}, whose generators take the form

\begin{eqnarray}
\ell _{n}^{\pm } &=&-\frac{1}{2}e^{\frac{n}{l}\left( t\mp li\phi \right)
}(l\partial _{t}-nr\partial _{r}\pm i\partial _{\phi })+\mathcal{O}(r^{-1})
\label{AKV} \\
&=&-ie^{\mp inx^{\pm }}(\pm \partial _{\pm }+inr\partial _{r})+\mathcal{O}%
(r^{-1}),
\end{eqnarray}%
where $x^{\pm }\equiv \phi \pm i{t}/{l}.$ For $n=0,\pm 1$, these coincide
with generators $J^+_{0}$ and $J^+_{\pm 1}$ (and the ones with minus superscript) at large $r$ up to order $\mathcal{O}%
(r^{-1})$. These vectors fields satisfy the conjugate relation $(\ell
_{n}^{+})^{\ast }=\ell _{n}^{-}$ and the Lie commutator $[\ell _{n}^{\pm },\ell
_{m}^{\pm }]=(n-m)\ell _{n+m}^{\pm }.$

Diffeomorphisms generated by $\ell^{\pm }_n$ preserve a set of boundary
conditions specified by their deviation with respect to the dS$_3$ metric (%
\ref{EldS}); namely $g_{\mu\nu}\to g_{\mu\nu}+\Delta g_{\mu\nu}$ with
\begin{eqnarray}  \label{BCdS}
\Delta g_{rr} &=& h_{rr}(x^+,x^-) r^{-3} + f_{rr}(x^+,x^-) r^{-4} + \cdots
\notag \\
\Delta g_{rj} &=& h_{rj}(x^+,x^-) r^{-2} + f_{rj}(x^+,x^-) r^{-3} + \cdots \\
\Delta g_{ij} &=& h_{ij}(x^+,x^-) r + f_{ij}(x^+,x^-) + \cdots  \notag
\end{eqnarray}
where $i,j \in \{+,-\}$.

Boundary conditions (\ref{BCdS}) are the $l\rightarrow il$ analytic
continuation of the relaxed AdS$_{3}$ boundary conditions proposed in \cite%
{OTT}, in which the $f_{ij}$, $f_{ir}$, and $f_{rr}$ components correspond
to the standard Brown-Henneaux fall-off conditions \cite{BH}. To study the
hairy black holes solutions (where $h_{ij}$, $h_{ir}$, and $h_{rr}$ are
non-vanishing), we need to include this relaxed set of boundary conditions.
In the coordinates employed to describe the dS$_{3}$ asymptotic conditions
in \cite{dSCFT},
the relaxed boundary conditions we will consider correspond to perturbing
the dS$_{3}$ metric $ds^{2}/l^{2}=e^{2t}dzd\bar{z}-dt^{2}$ with functions
obeying the asymptotic behavior $g_{z\bar{z}}={l^2}{e^{2t}}/{2}+{\mathcal{O}}%
(e^{{+}t})$, $g_{tt}=-l^2+{\mathcal{O}}(e^{{-}t})$, $g_{zz}={\mathcal{O}}%
(e^{t})$, ${g_{zt}= {\mathcal{O}}(e^{-2 t})}$ when $t\rightarrow {+}\infty $.

It is often convenient to work in Fefferman-Graham-like gauge, in which an
asymptotically dS$_3$ metric takes the form
\begin{equation}  \label{FG}
ds^2 = -\frac{l^2 d\rho^2}{\rho^2} + \underset{p \in \mathbb{N}}{\sum}
g_{ij}^{(p)} \rho^{2-p} dx^i dx^j
\end{equation}
where $g_{\pm \pm}^{(0)}=0$, $g_{+-}^{(0)}={l^2}/{2}$. In pure Einstein
gravity, the expansion terminates and the most general such solution is
known in closed form (see for instance Refs. \cite%
{1305.1277,Skenderis:2002wp}). In particular, the only non-vanishing metric
components are
\begin{eqnarray}  \label{dS3Gen}
g_{\pm \pm}^{(2)} = \frac{l}{2} L^\pm (x^\pm), \quad g_{\pm \pm}^{(4)} =
\frac{1}{4} L^+ L^-.
\end{eqnarray}
Note that because of the reality condition on $x^\pm$, the (complex)
functions $L^+$ and $L^-$ are related by $(L^+)^* = L^-$ in order for the
metric to be real.

In other gravity theories, on the other hand, an expansion in integer powers
$p$ is not the most general a priori. For instance, in the case of
asymptotically AdS spaces in BHT, it was argued in \cite{Cunliff} that this
expansion is not appropriate to capture all possible solutions, and one can
expect a similar situation in dS space. For the solutions we will be
interested in, however, it will be sufficient to consider $p\in \mathbb{N}$.
In BHT for generic values of the couplings, the equations of motion impose
that the only non-vanishing components in the expansion (\ref{FG}) satisfy (%
\ref{dS3Gen}). At particular values, however, the situation changes.
Expanding the equations of motion as
\begin{equation}
E_{\mu \nu }=\underset{p\in \mathbb{N}}{\sum }E_{\mu \nu }^{(p)}\rho ^{2-p},
\end{equation}%
one finds for instance that
\begin{equation}
E_{ij}^{(1)}\sim g_{ij}^{(1)}(m^{2}l^{2}+\frac{1}{2}),\;E_{\rho \rho
}^{(6)}\sim g_{+-}^{(2)}(m^{2}l^{2}+\frac{1}{2}),\cdots
\end{equation}%
so that, as a consequence, the components $g_{ij}^{(1)}$, $g_{ij}^{(3)}$ and
$g_{+-}^{(2)}$ are no longer forced to vanish at the point $m^{2}l^{2}+1/2=0$%
. Other special points in the parameter space also appear \cite{Cunliff},
but here we will not focus on those.

Conserved charges in BHT theory can be obtained in the covariant formalism
by integrating the one-form potential
\begin{equation}
k_{\xi }[\delta g,g]\equiv\epsilon _{\phi \mu \nu }k_{\xi }^{[\mu \nu
]}[\delta g ,g]d\phi ,
\end{equation}
which depends on an asymptotic Killing vector $\xi $, the background metric $%
g$, and the linearized metric perturbation $\delta g$. Its explicit
expression, which depends only on the equations of motion, is what is called
$Q^{\mu\nu}(\xi)$ in Eq.~(9) and Eq.~(29) of Ref. \cite{otro}. The resulting
charge takes the form
\begin{equation}
Q_{\xi }=\int_{\bar{g}}^{g}d\delta g\oint_{S^{\infty }}k_{\xi }[\delta g,g]
\label{Qq}
\end{equation}%
where the inner integral integrates over the constant $t$ and $r$ slice at $%
r \rightarrow \infty$. 
This expression computes the infinitesimal charge difference between two
nearby solutions, $g$ and $g+\delta g$, while the outer integral is
performed in the space of parameters.

For the boundary conditions (\ref{BCdS}), the charges associated to the
asymptotic Killing vectors $\ell _{n}^{\pm }$ are%
\begin{eqnarray}
Q_{\ell _{n}^{+}} &=&\frac{1}{4\pi Gl}\int d\phi \;e^{-inx^{+}}(g_{++}^{(2)}-%
\frac{1}{l^{2}}g_{++}^{(1)}g_{+-}^{(1)})  \notag  \label{ChargesQ} \\
Q_{\ell _{n}^{-}} &=&\frac{1}{4\pi Gl}\int d\phi \;e^{inx^{-}}(g_{--}^{(2)}-%
\frac{1}{l^{2}}g_{--}^{(1)}g_{+-}^{(1)})
\end{eqnarray}

The non-linear part coming from the non-vanishing $g_{ij}^{(1)}$ metric
components has apparently been overlooked in earlier literature. The charges
are manifestly finite and integrable. They are also conserved by virtue of
the equations of motion
\begin{equation}
E_{r\pm}^{(5)} \sim \partial_{\mp} (g_{\pm \pm}^{(2)} - \frac{1}{l^2} g_{\pm
\pm}^{(1)} g_{+-}^{(1)}).
\end{equation}

The charges (\ref{ChargesQ}) generate through their Poisson bracket an
algebra isomorphic to that of the asymptotic Killing vectors, up to possible
central extension. There have been discussions in earlier literature about
the reality of such central extension in the dS$_{3}$ case\footnote{%
See for instance Refs. \cite{Park:1998qk, Balasubramanian, Marolf:2012kh}
and references therein.}, which depends on the hermiticity conditions
satisfied by the generators in relation to the inner product \cite%
{Bousso:2001mw,Witten:2001kn}. To understand this, let us first review the situation in AdS$_{3}$: In that
case, the asymptotic Killing vectors $\tilde{\ell}_{n}$ satisfy
\begin{equation}
i[\tilde{\ell}_{n}^{\pm },\tilde{\ell}_{m}^{\pm }]=(n-m)\tilde{\ell}%
_{n+m}^{\pm }  \label{alg2}
\end{equation}%
and the hermiticity condition
\begin{equation}
(\tilde{\ell}_{n}^{\pm })^{\ast }=\tilde{\ell}_{-n}^{\pm }.  \label{conj2}
\end{equation}

Then, when realized as asymptotically conserved charges $Q_{\tilde{\ell}%
_{n}^{\pm }}$, the algebra they generate is represented by a covariant
Poisson bracket \cite{BB}, which (up to field redefinitions) reads
\begin{equation}
\{Q_{\tilde{\ell}^\pm_{n}},Q_{\tilde{\ell}^\pm_{m}}\}\equiv \delta _{\tilde{%
\ell}^\pm_{m}}Q_{\tilde{\ell}^\pm_{n}}=Q_{[\tilde{\ell}^\pm_{n},\tilde{\ell}%
^\pm_{m}]}+K_{\tilde{\ell}^\pm_{n},\tilde{\ell}^\pm_{m}},
\end{equation} where the term $K_{\tilde{\ell}^\pm_{n},\tilde{\ell}^\pm_{m}}$ corresponds
to a possible central extension, given by
\begin{equation}
K_{\tilde{\ell}^\pm_{n},\tilde{\ell}^\pm_{m}}=\oint_{S^{\infty }}k_{\tilde{%
\ell}^\pm_{n}}[\mathcal{L}_{\tilde{\ell}^\pm_{m}}\bar{g},\bar{g}].
\end{equation}%}
This quantity is imaginary for AdS$_{3}$. Now, replacing Poisson brackets by
commutators and charges by quantum operators, $i\{Q_{\tilde{\ell}%
^\pm_{m}},Q_{\tilde{\ell}^\pm_{n}}\}\rightarrow \lbrack \tilde{L}^\pm_{m},%
\tilde{L}^\pm_{n}]$, leads to the expected Virasoro algebra
\begin{equation}
\lbrack \tilde{L}^\pm_{n},\tilde{L}^\pm_{m}]=(n-m)\tilde{L}^\pm_{n+m}+\frac{%
c_{\text{AdS}}}{12}(n^{3}-n)\delta _{m+n}  \label{Vir}
\end{equation}%
with $c_{\text{AdS}}$ being the Brown-Henneaux central charge \cite{BH}. The
hermiticity condition of the corresponding quantum operators is the standard
relation
\begin{equation}
(\tilde{L}_{n}^{\pm })^{\dagger }=\tilde{L}_{-n}^{\pm },  \label{hermi}
\end{equation}%
which follows from hermiticity of the stress-tensor.

In the dS$_{3}$ case, in contrast, we have seen that the asymptotic Killing
vectors satisfy the Lie commutator
\begin{equation}
\lbrack \ell _{n}^{\pm },\ell _{m}^{\pm }]=(n-m)\ell _{n+m}^{\pm }
\label{alg3}
\end{equation}%
with the condition
\begin{equation}
(\ell _{n}^{\pm })^{\ast }=\ell _{n}^{\mp },  \label{conj3}
\end{equation}%
which differs from (\ref{conj2}). In the context of de Sitter space, on the
other hand, the question arises as to what is the natural adjoint relation to
impose on the quantum generators. Such questions have been discussed in
various works \cite%
{Witten:2001kn,Balasubramanian,Bousso:2001mw,Ng:2012xp,Jafferis:2013qia}. In
particular, in Ref. \cite{Bousso:2001mw} the different hermiticity
conditions are discussed in relation to the different manners of defining an
inner product on dS \cite{Witten:2001kn}. Here, following \cite%
{Bousso:2001mw,Ng:2012xp}, we adopt the hermiticity condition (\ref{hermi})
for $L_n^\pm$. Consistency with (\ref{alg3})-(\ref{conj3}) then requires that the
quantization rule now becomes\footnote{%
More precisely, consider $\alpha \{Q_{\ell _{n}^{\pm }},Q_{\ell _{m}^{\pm
}}\}\equiv \delta _{\ell _{n}^{\pm }}Q_{\ell _{m}^{\pm }}=Q_{[\ell _{n}^{\pm
},\ell _{m}^{\pm }]}+K_{\ell _{n}^{\pm },\ell _{m}^{\pm }}$, and
quantization rule $\beta \{Q,Q\}\rightarrow \lbrack L,L]$. We have to
determine $\alpha $ and $\beta $. Now, taking the complex conjugate of the Poisson bracket $\alpha \{Q_{\ell
_{n}^{\pm }},Q_{\ell _{m}^{\pm }}\}$ and using the fact that $Q_{\ell
_{n}^{+}}^{\ast }=Q_{\ell _{n}^{-}}$, we deduce that $\alpha $ is real.
On the other hand, taking the dagger of $\alpha \lbrack
L_{n}^{\pm},L_{m}^{\pm}]$ and using $(L_{n}^{\pm })^{\dagger }=L_{-n}^{\pm }$
we conclude that both $\beta $ and $K_{\ell _{n}^{\pm },\ell _{m}^{\pm }}$
are real. This is what led to modifying the quantization rule. We thank G.
Comp\`{e}re for helpful discussion on this point.} $\{Q_{\ell^\pm
_{m}},Q_{\ell^\pm _{n}}\}\rightarrow \lbrack L^\pm_{m},L^\pm_{n}]$.

The unusual quantization rule can be motivated heuristically by thinking in the
example of a harmonic oscillator: There, the quantization rule $%
i\{.,.\}\rightarrow [.,.]$ is equivalent to the reality condition on the
position (and its conjugate momentum) and the requirement that the
corresponding operators to be hermitian. If one formally Euclideanizes time,
the corresponding Euclidean momentum $p_{E}\equiv ip$ will satisfy $%
p_{E}^{\ast }=-p_{E}$ (complex conjugate acts as Euclidean time reflection),
implying that the corresponding quantum operator (for the Euclidean
momentum) is naturally anti-hermitian rather than hermitian. This leads to a
modified quantization rule of the form $\{.,.\}\rightarrow \lbrack .,.]$. In
the present context, we would like to suggest that a similar rule is perhaps
sensible in de Sitter gravity where the static time outside the horizon is
Euclidean\footnote{%
As expressed in \cite{Balasubramanian}, relating a Lorentzian bulk with an
Euclidean boundary introduces some extra factors of $i$ and unusual reality
conditions which still need to be understood better in order to formulate a
correspondence such as dS/CFT.}. If one adopts such suggestion, the algebra
of quantum operators turns out to be
\begin{equation}
\lbrack L_{n}^{\pm },L_{m}^{\pm }]=(n-m)L_{n+m}^{\pm }+K_{\ell _{n}^{\pm
},\ell _{m}^{\pm }},
\end{equation}%
where
\begin{equation}
K_{\ell _{n}^{\pm },\ell _{m}^{\pm }}=\frac{c^{\pm }}{12}(n^{3}-n
)\delta _{m+n},  \label{eq:centralchargeCanonical}
\end{equation}%
with the central charge given by
\begin{equation}
c^{\pm }=-\frac{3l}{2G}\left( 1-\frac{1}{2m^{2}l^{2}}\right) .  \label{Thec}
\end{equation}

We observe that, as it happens in the AdS$_{3}$ case, central charge (\ref%
{Thec}) receives a contribution $\sim 1/m^{2}$ coming from the
higher-curvature terms. In the limit $m^{2}\rightarrow \infty $ one recovers
the central charge of \cite{Balasubramanian} which differs from \cite{dSCFT}
by a minus sign.

In the case of $m^2 < 0$ (which is what we are interested here), the central charge (\ref{Thec}) is negative. However, one
can always change the sign of $c^{\pm }$ by redefining Virasoro generators
as $\ell _{n}^{\pm }\rightarrow -\ell _{-n}^{\pm }$. However, this would map
an spectrum for the $(L_0^+ + L_0^-)$ operator that is bounded from below --
see (\ref{deBuyl}), (\ref{40bis}) and (\ref{50}) below -- into a spectrum
that is bounded from above. A convention-independent question is to ask if
$c (L_0^+ + L_0^-)$ is bounded below or above. A usual unitary CFT would
have $c (L_0^+ + L_0^-)$ bounded below, while in our case this is bounded
above.

Notice that a naive Euclidean continuation from AdS$_{3}$ to dS$_{3}$ would
have led to an imaginary central charge, as pointed out for instance in \cite%
{Maldacena:2002vr,Balasubramanian,Park:1998qk,Marolf}. This differs from the
real central charge obtained above as well as from the result from
Brown-York stress-tensor to be discussed in Sec. (\ref{BY}). Such difference
has to do with the different choice of hermiticity (and hence quantization
rule) which is closely related to the convention of stress tensor. One
could, alternatively, consider a imaginary central charge $ic^{\pm }$ by
appropriately redefining Virasoro generators and try to make sense out of CFT%
$_{2}$ computation we will discuss in Section IV. Here we will work with a
real central charge, which is negative at the special point (\ref{dD}). This
is what is expected if one closely follows the relationship to AdS$_{3}$.
Indeed, the Virasoro generators in that case read
\begin{equation}
\tilde{\ell}_{n}^{\pm }=\frac{1}{2}e^{in\left( t/l_{\text{AdS}}\pm \phi
\right) }(l_{\text{AdS}}\partial _{t}-inr\partial _{r}\pm \partial _{\phi })+%
\mathcal{O}(r^{-1}),  \label{AKVAdS}
\end{equation}%
where $l_{\text{AdS}}$ is the AdS radius.
At the level of Poisson bracket, it is easy to see that under analytic
continuation $l_{\text{AdS}}\rightarrow -il,n\rightarrow -n$, the generators
map into $\tilde{\ell}_{n}^{\pm }\rightarrow i\ell _{n}^{\pm }$, and hence
the Poisson bracket of the dS$_{3}$ charges satisfy
\begin{equation}
\{Q_{\ell^\pm_{n}},Q_{\ell^\pm_{m}}\}=(n-m)Q_{\ell^\pm_{m+n}}+\frac{c^{\pm}}{12}
(n^{3}-n)\delta _{n+m},
\end{equation} with $c^{\pm}$ given in (\ref{Thec}).
Hence, the central charge is negative provided one keeps the usual Virasoro
structure constants.

\section{Black holes in de Sitter space}

\label{sec:dSBH}

\subsection{Static black holes}

At the partially massless point (\ref{dD}), the theory (\ref{SW}) exhibits
special properties as we observed on general grounds in the previous
section. In particular, at that point the following metric solves the
equations of motion (\ref{fieldeq})%
\begin{equation}
ds^{2}=-\frac{(r-r_{+})(r_{++}-r)}{l^{2}}dt^{2}+\frac{l^{2}dr^{2}}{%
(r-r_{+})(r_{++}-r)}+r^{2}d\phi ^{2},  \label{uds}
\end{equation}%
where $r_{+}$ and $r_{++}$ are two integration constants.

The black hole solution (\ref{uds}) was first derived in Ref. \cite{NMG2,
OTT}, and it can be shown to be the most general spherically symmetric
solution to the equations of motion of BHT theory in asymptotically dS
space. Provided that both $r_{+}$ and $r_{++}$ are positive, $r_{+}$ represents the location of the black hole event horizon while $r_{++}\ge
r_{+}$ is the cosmological horizon. It is convenient to define the
parameters $b$ and $M$ by $r_{++}+r_{+}\equiv bl^{2}$, $r_{++}r_{+}\equiv
4MGl^{2}$. They will be related to the black hole mass (see (\ref{deBuyl})
below). Empty dS$_{3}$ space corresponds to the particular case $%
r_{++}=-r_{+}=\pm l$, while $r_{++}=-r_{+}\neq \pm l$ correspond to
non-rotating Kerr-dS$_{3}$ solutions (see Refs. \cite{Park:1998qk,
Balasubramanian, Bousso:2001mw} for discussions of the latter solution in
the context of dS/CFT). For general $r_{++}\neq -r_{+}$, there is a
curvature singularity at $r=0$ as can be seen from the Ricci scalar $%
R=6/l^{2}-2b/r$. In dS space, one would like the observer between $r_{+}$
and $r_{++}$ not to be exposed to a naked singularity\footnote{%
For Kerr-dS$_{3}$ in pure gravity, however, a naked conical defect is fine
since it corresponds to some point-like or small massive object at the
origin.}, since this observer is the natural static observer (where the
Killing vector $\partial _{t}$ is timelike). This means that we should
insist that $r_+$ is real and non-negative. This gives rise to the
inequalities $b^{2}l^{2}\ge 16GM \ge 0$. The Penrose diagram of the dS$_{3}$ black hole is identical to that of its
higher-dimensional analogues.

Using (\ref{Qq}) one can compute the zero-mode charges $L_{0}^{+}\equiv
Q_{\ell _{0}^{+}}$ and ${L}_{0}^{-}\equiv Q_{{\ell }_{0}^{-}}$, associated
to the Killing vectors $\ell _{0}^{\pm }=-\frac{1}{2}(l\partial _{t}\pm
i\partial _{\phi })$.
The conserved charge $Q_{\partial _{t}}\equiv {\mathcal{M}}$ associated to
solution (\ref{uds}) is\footnote{%
Note that the {\itshape black hole mass} is given by $-{\mathcal{M}}%
=-Q_{\partial_t}$, cf. Eq. (2.8) of \cite{Anninos:2010gh}.} 
\begin{equation}
\mathcal{M}=-\frac{1}{16Gl^{2}}(r_{++}-r_{+})^{2}=M-\frac{b^{2}l^{2}}{16G}.
\label{deBuyl}
\end{equation}

Now, let us consider the dS$_{3}$ metric, which corresponds to $%
r_{++}=-r_{+}=l$. In this case we obtain the energy difference between dS$%
_{3}$ and the massless (${\mathcal{M}}={\mathcal{J}}=0$) solution\footnote{%
Notice that there is a one-parameter family of solutions with $\mathcal{M}={\mathcal{J}}=0$.
This is reminiscent of what happens in TMG\ at the chiral point, where all
BTZ\ solutions with $M_{BTZ}=J_{BTZ}$ have vanishing charges.}
\begin{equation}
\Delta E\equiv \left. Q_{\partial _{t}}\right\vert _{dS_{3}}=-\frac{1}{4G}.
\label{eldentes}
\end{equation}%
Notice this is consistent with
\begin{equation}
\frac{c^{\pm }}{12}=-\frac{l}{4G}
\end{equation}%
according to the CFT$_{2}$ interpretation, due the identification
\begin{equation}
L_{0}^{+}+L_{0}^{-}=-lQ_{\partial _{t}}.  \label{40bis}
\end{equation}

\subsection{Rotating black holes in dS}

One may also provide the black hole (\ref{uds}) with angular momentum. In
that case the metric acquires an abstruse form: In a system that asymptote dS%
$_{3}$ in its form (\ref{EldS}) the rotating black hole solution reads%
\begin{equation}
ds^{2}=-N^{2}(r)F(r)dt^{2}+\frac{dr^{2}}{F(r)}+r^{2}\left( d\phi +N^{\phi
}(r)dt\right) ^{2}\ ,  \label{Hair}
\end{equation}%
where $N(r)$, $N^{\phi }(r)$, and $F(r)$ are functions of the coordinate $r$%
, given by%
\begin{equation}
F(r)=\frac{\hat{r}^{2}}{r^{2}}\left[ -\frac{\hat{r}^{2}}{l^{2}}+\frac{b}{2}%
\left( 1+\eta \right) \hat{r}-\frac{b^{2}l^{2}}{16}\left( 1-\eta \right)
^{2}-4MG\eta \right]
\end{equation}%
and%
\begin{equation}
N(r)=1-\frac{bl^{2}}{4\hat{r}}\left( 1-\eta \right) ,\quad N^{\phi }(r)=-%
\frac{a}{2r^{2}}\left( 4GM-b\hat{r}\right) ,  \label{HH}
\end{equation}%
where $\hat{r}^{2}=r^{2}+2MGl^{2}\left( 1-\eta \right)
-(b^{2}l^{4}/16)\left( 1-\eta \right) ^{2}$ and where $\eta =\pm \sqrt{%
1+a^{2}/l^{2}}$ is a rotation parameter. Again, these black holes exist only
at the partially massless point $m^{2}l_{-}^{2}=m^{2}l_{+}^{2}=-1/2$ of the
parameter space.

For convenience, we will work with $\hat{r}$ being the radial
coordinate where the metric reads%
\begin{equation}
ds^{2}=-N^{2}(\hat{r})F(\hat{r})dt^{2}+\frac{\hat{r}^{2}d\hat{r}^{2}}{[r(%
\hat{r})]^{2}F(\hat{r})}+[r(\hat{r})]^{2}\left( d\phi +N^{\phi }(\hat{r}%
)dt\right) ^{2}.  \label{HairRhat}
\end{equation}%
We will work with $b\geq 0$, since for the opposite sign we could redefine $%
\hat{r}\rightarrow -\hat{r}$ to absorb the sign. 
One can see that there is a curvature singularity at
\begin{equation}
\hat{r}=\hat{r}_{s}\equiv \frac{b}{4}(1-\eta )l^{2},
\end{equation}%
as can be seen from the Ricci scalar
\begin{equation}
R=\frac{6}{l^{2}}-\frac{2b\eta a}{\hat{r}-\hat{r}_{s}}.
\end{equation}%
The horizons are now located at
\begin{eqnarray}
\hat{r}_{++} &=&\frac{1}{4}b(1+\eta )l^{2}+2l\sqrt{-G\eta \left( M-\frac{%
b^{2}l^{2}}{16G}\right) },  \notag \\
\hat{r}_{+} &=&\frac{1}{4}b(1+\eta )l^{2}-2l\sqrt{-G\eta \left( M-\frac{%
b^{2}l^{2}}{16G}\right) }.
\end{eqnarray}%
If the two horizons are shielding the singularity, i.e. $\hat{r}_{++}\ge
\hat{r}_{+}\ge \hat{r}_{s}$, then we have 
$\eta \ge1, \hat{r}_{s} \le 0$ and 
\begin{equation}
0\leq-\left(M-\frac{b^2 l^2}{16 G}\right)\le \frac{b^{2}\eta l^{2}}{16G}.
\end{equation}%
In short, we will focus on the following case:
\begin{equation}
b\geq 0\quad ,\quad \eta \geq 1\quad ,\quad 0\leq -\left(M-\frac{b^2 l^2}{16
G}\right)\le \frac{b^{2}\eta l^{2}}{16G}.  \label{50}
\end{equation}

Introducing $\hat{r}_\phi \equiv 4GM/b$ and $\hat{r}_0^2\equiv2(1-\eta)G l^2
M -b^2l^4(\eta-1)^2/16 =2(1-\eta)G l^2 M -\hat{r}_s^2 $, we can rewrite the
metric in a nicer form
\begin{eqnarray}  \label{nice:metric}
ds^2 &=& -\frac{\left(\hat{r}-\hat{r}_s \right)^2\left(\hat{r}_{++}-\hat{r}%
\right) \left(\hat{r}-\hat{r}_+ \right)}{l^2\left(\hat{r}^2-\hat{r}%
_0^2\right)} dt^2  \notag \\
& &+\frac{l^2 d\hat{r}^2}{\left(\hat{r}_{++}-\hat{r}\right)\left( \hat{r}-%
\hat{r}_+\right)}  \notag \\
& &+ \left( \hat{r}^2-\hat{r}_0^2\right)\left[ d\phi+\frac{a b \left(\hat{r}-%
\hat{r}_\phi\right)}{2(\hat{r}^2-\hat{r}^2_0)}dt \right]^2.
\end{eqnarray}
Note that the function $N(r)$ now becomes
\begin{equation}
N(\hat{r})=1- \frac{\hat{r}_s}{\hat{r}},
\end{equation}
and we have $\sqrt{-g}=(\hat{r}-\hat{r}_s)$ while the determinant of the
induced $(t,\phi)$-metric is $(\hat{r}-\hat{r}_{++})(\hat{r}-r_+)(\hat{r}-%
\hat{r}_s)^2/l^2$.

The charges (\ref{Qq}) of the rotating solution (\ref{Hair})-(\ref{HH}) are
\begin{eqnarray}
\mathcal{M} &\equiv& Q_{\partial_t} =M-\frac{b^{2}l^{2}}{16G} =-\frac{1}{16
G l^2 \eta}\left(\hat{r}_{++}-\hat{r}_{+}\right)^2 ,  \notag \\
\mathcal{J}&\equiv&Q_{\partial_\phi}=J+a\frac{b^{2}l^{2}}{16G} ,
\label{carguitas}
\end{eqnarray}
where $J \equiv -a M$ and $\mathcal{J} \equiv -a \mathcal{M}$. 

\subsection{Boundary stress-tensor}

\label{BY}

In this section, we would like to show that the results for the conserved
charges in previous sections are in agreement with the proposal of Ref.~\cite%
{Balasubramanian,Skenderis:2002wp,dSCFT2} where a prescription to define a
stress-tensor associated to the boundary dual CFT$_{2}$ in the context of
dS/CFT was given. The idea in \cite{Balasubramanian,Skenderis:2002wp,dSCFT2}
is adapting the holographic renormalization recipe of AdS/CFT \cite%
{Balasubramanian:1999re,de Haro:2000xn,Skenderis:2002wp} to the case of
asymptotically dS spaces. This consists in defining a regularized version of
the Brown-York tensor at the boundary. In the case of asymptotically dS
spaces, we will focus on the boundary corresponds to the Euclidean surfaces
at future null infinity $\mathcal{I}^{+}$

As in the case of AdS, to apply the method of \cite%
{Balasubramanian,Skenderis:2002wp,dSCFT2} one first needs to supplement the
gravity action with adequate boundary terms, which eventually yield the
correct boundary stress-tensor. In order to do so we will follow the
prescription given in \cite{HohmTonni}, which amounts to first rewriting BHT
action in an alternative way. That is, consider the action%
\begin{eqnarray}
S_{\text{BHT}} &=&\frac{1}{16\pi G}\int_{\Sigma }{d^{3}x\sqrt{-g}\left(
f^{\mu \nu }(R_{\mu \nu }-\frac{1}{2}Rg_{\mu \nu })-\right. }  \notag \\
&&{\frac{1}{4}m^{2}(f_{\mu \nu }f^{\mu \nu }-f^{2}))},  \label{SNMG}
\end{eqnarray}%
which, apart from the metric, involves an auxiliary symmetric rank-two field
$f_{\mu \nu }$. Being non-dynamical, $f_{\mu \nu }$ field can be integrated
and the result plugged back into (\ref{SNMG}). This can be easily shown to
reproduce the original BHT action. However, the alternative form for the
action (\ref{SNMG}) results convenient to write down the boundary terms.

According to the prescription in \cite{HohmTonni}, boundary terms, $S_{\text{%
B}}$, are to be introduced in the action for the variational principle to be
defined in such a way that both the metric $g_{\mu \nu }$ and the auxiliary
field $f_{\mu \nu }$ are fixed on the boundary $\partial \Sigma $. Here, we study asymptotically dS in the future, where the boundary corresponds to future null infinity $\mathcal{I}^{+ }$.

With this prescription, the boundary action $S_{\text{B}}$ reads%
\begin{equation}
S_{\text{B}}=-\frac{1}{8\pi G}\int_{\mathcal{I}^+ }d^{2}x\sqrt{|\gamma |}%
\left( K+\frac{1}{2}\hat{f}^{ij}(K_{ij}-\gamma _{ij}K)\right) ,  \label{Sb}
\end{equation}
where $\gamma _{ij}$ are the components of the induced metric, $\gamma $ its
determinant, and $K_{ij}$ is the extrinsic curvature on $\mathcal{I}^+ $. We
shall define $\hat{f}^{ij}$ below.

The induced metric $\gamma _{ij}$ of the boundary has components $i,j\in
\{t,\phi \}$ and is Euclidean.
Note that $\partial _{t}$ is timelike inside the static
patch (i.e. inside the cosmological horizon seen by an inertial observer in
dS space) while it is spacelike outside the static patch. In particular, it
is spacelike at $\mathcal{I}^{+}$ which is at large $r\gg l$; see \cite%
{Balasubramanian}.
One considers the following
ADM\ like coordinates%
\begin{equation}
ds^{2}=-N^{2}dr^{2}+\gamma _{ij}(dx^{i}+N^{i}dr)(dx^{j}+N^{j}dr),
\end{equation}%
where $N^{2}$ is the radial lapse function, and $\gamma _{ij}$ is then
defined on constant-$r$ surfaces.

In terms of these coordinates, field $\hat{f}^{ij}$ appearing in (\ref{Sb})\
comes from decomposing the contravariant auxiliary field $f^{\mu \nu }$ as%
\begin{equation}
f^{\mu \nu }=\left(
\begin{array}{cc}
f^{ij} & h^{j} \\
h^{i} & s%
\end{array}%
\right)
\end{equation}%
and then defining
\begin{equation}
\hat{f}^{ij}\equiv f^{ij}+2h^{(i}N^{j)}+sN^{i}N^{j}.
\end{equation}

Then, the Brown-York stress-tensor is defined by varying\footnote{%
In this auxiliary field formalism, the appropriate independent field at the
boundary is $f_i{}^j$, i.e. the variation of $\gamma^{ij}$ acts trivially on
$f_i{}^j$. For more details, see \cite{HohmTonni}.} the action $S_{\text{A}%
}+S_{\text{B}}$ with respect to the metric $\gamma ^{ij}$. That is,%
\begin{equation}
T_{ij}=\left. \frac{2}{\sqrt{|\gamma |}}\frac{\delta S}{\delta \gamma ^{ij}}
\right|_{r=\text{const}}.  \label{Tij}
\end{equation}

This yields $T^{ij}=T_{\text{GR}}^{ij}+T_{\text{BHT}}^{ij}$, which consists
of the standard Israel contribution%
\begin{equation}
T_{\text{GR}}^{ij}=\frac{1}{8\pi G}(K^{ij}-K\gamma ^{ij}),
\end{equation}%
supplemented by a contribution coming from the higher-curvature terms \cite%
{HohmTonni}%
\begin{eqnarray}
T_{\text{BHT}}^{ij} &=&-\frac{1}{8\pi G}\left( \frac{1}{2}\hat{f}%
K^{ij}+\nabla ^{(i}\hat{h}^{j)}-\frac{1}{2}D_{r}\hat{f}^{ij}+K_{k}^{(i}\hat{f%
}^{j)k}\right. -  \notag \\
&&\left. \frac{1}{2}\hat{s}K^{ij}-\gamma ^{ij}(\nabla _{k}\hat{h}^{k}-\frac{1%
}{2}\hat{s}K+\frac{1}{2}\hat{f}K-\frac{1}{2}D_{r}\hat{f})\right) ,  \notag \\
&&
\end{eqnarray}%
where $\hat{h^{i}}=N(h^{i}+sN^{i}N^{j})$, $\hat{s}=-N^{2}s$, and $\hat{f}%
\equiv \gamma _{ij}\hat{f}^{ij}$. The covariant $r$-derivative $D_{r}$ is
defined as
\begin{eqnarray}
D_{r}\hat{f}^{ij} &\equiv &-\frac{1}{N}\left( \partial _{r}\hat{f}%
^{ij}-N^{k}\partial _{k}\hat{f}^{ij}+2\hat{f}^{k(i}\partial
_{k}N^{j)}\right) ,  \notag \\
D_{r}\hat{f} &\equiv &-\frac{1}{N}\left( \partial _{r}\hat{f}-N^{k}\partial
_{k}\hat{f}\right) .
\end{eqnarray}

In order to define the stress-tensor at the boundary of the space it is
necessary to regularize the divergence that appears in large $r$ limit. As
in the case of asymptotically AdS$_{3}$ spaces, this is achieved by adding
boundary terms constructed out of intrinsic boundary quantities. Such terms
have the generic form
\begin{equation}
S_{\text{C}}=\frac{1}{8\pi G}\int_{\mathcal{I}^+ }d^{2}x\sqrt{|\gamma |}
(\alpha _{0}+\alpha _{1}\ \hat{f}+\alpha _{2}\ \hat{f}^{2}+\beta _{2}\ \hat{f%
}_{ij}\hat{f}^{ij}+...)  \label{Sc}
\end{equation}%
where $\alpha _{i},$ $\beta _{i}$, etc are constant coefficients.

The regularized boundary stress-tensor is then defined by taking the $%
r\rightarrow \infty $ limit of the improved stress-tensor%
\begin{equation}
T_{ij}\rightarrow T_{ij}^{(\text{reg})}=T_{ij}+\frac{2}{\sqrt{|\gamma |}}
\frac{\delta S_{\text{C}}}{\delta \gamma ^{ij}}.  \label{Tijren}
\end{equation}

{\ In the present case of $m^2 l^2=-1/2$, where the hairy black holes exists,%
} the choice of counterterms (\ref{Sc}) that makes the action (and both the mass and angular momentum) finite is $\alpha _{1}=-1/l,$ with
all the other coefficients $\alpha _{i\neq 1}$, $\beta _{i}$, etc set to
zero. Since for the case of AdS$_3$ hairy rotating black holes (as discussed
in \cite{GiribetLeston}) one prescribes $\alpha_1=1/l_{\text{AdS}}$, the
above result for dS$_3$ is consistent with generalizing the prescription
discussed in \cite{Skenderis:2002wp} where the counter-terms in dS are the
same as that in AdS but with coefficients of opposite signs.

With this regularized stress-tensor defined at the boundary one proposes the
following definition of conserved charges%
\begin{equation}
Q_{\xi }=\int ds\ u^{i}T_{ij}^{(\text{reg})}\xi ^{j},  \label{carga}
\end{equation}%
where $ds$ is the line element of the constant-$t$ surfaces at $\partial
\Sigma $ (recall $t$ is spacelike at the boundary), $u$ is a unit vector
orthogonal to the constant-$t$ surfaces, and $\xi $ is the Killing vector
that generates the isometry on $\mathcal{I}^+ $.

As a result, one finds that the conserved charges of the rotating solution (\ref{Hair})-(\ref{HH}) are
\begin{equation}
Q_{\partial _{t}}=M-\frac{b^{2}l^{2}}{16G},\qquad Q_{\partial _{\phi }}=J+a%
\frac{b^{2}l^{2}}{16G},  \label{MandJ}
\end{equation}%
which exactly match (\ref{carguitas}).

{Given the renormalized stress-tensor, one can also follow the holographic
renormalization procedure\footnote{%
For such computations in the context of dS/CFT in pure gravity, see \cite%
{Balasubramanian,dSCFT2}.} of \cite{Balasubramanian:1999re,de
Haro:2000xn,Skenderis:2002wp} and derive the central charge from the
Schwarzian term in the transformation of the (renormalized) stress tensor.
For $m^2 l^2=-1/2$, this yields $c^{\pm }={3l}/{G}$ which, up to a sign, agrees with Eq.~(\ref{Thec}). This suggests to define the boundary stress-tensor as $%
T_{ij}\rightarrow -T_{ij}$ in (\ref{Tij}), and thus choosing the integration
orientation in (\ref{carga}) accordingly.}

\section{Thermodynamics}

\subsection{The dS$_{3}$ black hole thermodynamics}

Let us now discuss the black hole thermodynamics in dS$_{3}$ space, which is
actually our principal motivation. Consider first the case of static black
hole. The temperature of both the cosmological horizon and of the black hole
horizon computed from Eq.~(\ref{uds}) give%
\begin{equation}
T_{+}=T_{++}=\frac{1}{4\pi l^{2}}(r_{++}-r_{+}).  \label{rmkabl1}
\end{equation}%
That is, the dS black holes of three-dimensional BHT massive gravity are
always in thermal equilibrium with dS space that hosts them. Moreover, by
evaluating the Wald formula one finds that the Bekenstein-Hawking entropy of
the black hole $S_{+}$ coincides with the Gibbons-Hawking entropy of the
cosmological horizon $S_{++}$ given by 
\begin{equation}
S_{++}=S_{+}=\frac{\pi }{2G}(r_{++}-r_{+}).  \label{rmkabl2}
\end{equation}

Now, consider the case of rotating solution (\ref{Hair})-(\ref{HH}). The
temperatures associated to the horizons $\hat{r}_{+}$ and $\hat{r}_{++}$
read
\begin{equation}
T_{++}=T_{+}=\frac{\eta }{\pi l}\sqrt{\frac{-2G\mathcal{M}}{1+\eta }}=\frac{%
\gamma ^{-1}}{4\pi l^{2}}\left( \hat{r}_{++}-\hat{r}_{+}\right) ,
\label{Tmm}
\end{equation}%
where $\gamma \equiv \sqrt{(1+1/\eta )/2}$. As in the case of the static
black hole (where $\eta =1$), the Hawking temperature of the black hole
event, $T_{+} $, coincides with the Gibbons-Hawking temperature of the
cosmological horizon, $T_{++}$. The entropy associated to the horizon is
given by Wald formula which evaluates to be%
\begin{equation}
S_{++}=S_{+}=\pi l\sqrt{-\frac{2\mathcal{M}}{G}(1+\eta )}=\frac{\pi }{2G}%
\gamma \left( \hat{r}_{++}-\hat{r}_{+}\right) .  \label{astec}
\end{equation}

From the temperature and entropy in Eq.~(\ref{Tmm})-(\ref{astec}), together
with the conserved charges in Eq.~(\ref{carguitas}), one observes that the
first law
\begin{equation}
-d\mathcal{M}=T_{++}dS_{++}+\Omega _{++}d\mathcal{J},
\end{equation}%
is satisfied, where
\begin{equation}
\Omega _{++}=\Omega _{+}=-\frac{1}{l}\sqrt{\frac{\eta -1}{\eta +1}}.
\end{equation}%
The $r_{+}$ horizon then satisfies exactly the same relation. Let us define $%
E_{BH}\equiv -\mathcal{M}=-Q_{\partial _{t}}$, which is non-negative. Then
the black hole horizon satisfies the usual first law
\begin{equation}
dE_{BH}=T_{+}dS_{+}+\Omega _{+}d\mathcal{J}.
\end{equation}%
This definition of the energy of the black hole in dS (in particular the
sign) is already used for instance in Ref.~\cite{Dio,Anninos:2010gh} for
Kerr-dS$_{4}$ black holes so that the thermodynamics and charges of the
black hole horizon reduces to the standard one in flat space.

Note also that, from \cite{OTT}, we know that there are simple relations
between the thermodynamical quantities of the rotating black holes and those
of the non-rotating ones. In our convention, we actually have%
\begin{equation}
T_{++}=\gamma ^{-1}\sqrt{\eta }\times \left( \left. T_{++}\right\vert _{\eta
=1}\right) ,\quad S_{++}=\gamma \sqrt{\eta }\times \left( \left.
S_{++}\right\vert _{\eta =1}\right) .
\end{equation}%

An alternative way of writing the entropy is in terms of thermodynamical
quantities:
\begin{equation}
S_{++}=S_{+}=\frac{2l^{2}\pi ^{2}T_{++}}{G(1+l^{2}\Omega _{++}^{2})}.
\label{eq:entropyThermo}
\end{equation}

\subsection{Extremal limit}
\label{sec:ext}
Now, let us discuss the extremal limits of the dS$_{3}$ black hole solutions
and their thermodynamics. But let us first consider the rotating AdS$_{3}$ hairy black holes. They actually have two
extremal limits: the extremal rotating limit (like that in Kerr/CFT) and the
zero-entropy limit. In fact, in the Kerr/CFT limit, we have self-dual AdS$%
_{3}$ while in the zero entropy limit, we get AdS$_{2}\times S^{1}$. The extremal
rotating limit (corresponding to $\eta =0$) has no analogue for dS since $%
\eta \ge 1$. 

Let us focus back to the dS black holes. The strictly zero entropy limit ($%
r_+=r_{++}$) corresponds to $\mathcal{M}=0$ and has a near horizon geometry which is dS$_2 \times S^1$, already observed
in \cite{OTT} from Euclidean continuation of the instanton $S^2\times S^1$.

Alternatively, one can consider the Nariai-type limit\cite%
{Dio,Anninos:2010gh} of the metric in Eq.~(\ref{nice:metric}). First, define
the dimensionless near-extremal parameter $\lambda$
\begin{equation}
\lambda \equiv \frac{\hat{r}_{++}-\hat{r}_+}{\hat{r}_{++}},
\end{equation}
which is proportional to the temperature of the horizon near extremality
\begin{equation}
T_{++}=\frac{b (1+\eta)}{16\pi \gamma}\lambda +\mathcal{O}(\lambda^2).
\end{equation}
Now, perform the scaling
\begin{equation}  \label{nariai:limit}
\tau=\epsilon k t,\quad x=\frac{\hat{r}_{++}-\hat{r}}{\hat{r}_{++}\epsilon }%
,\quad \tilde{\phi}=\phi- \Omega_+ t
\end{equation}
taking $\epsilon\rightarrow0$ with $\lambda/\epsilon$ kept finite, one gets
the near-horizon metric 
\begin{equation}
ds^2=-\frac{x(\lambda-x)}{l^2}d\tau^2 +\frac{l^2 dx^2}{x(\lambda-x)} + k^2 d%
\tilde{\phi}^2, \quad k^2\equiv \hat{r}_{++}^2-\hat{r} _0^2.
\end{equation}
Note that this is just dS$_2\times S^1$ where the dS$_2$ temperature is
\begin{equation}
T_{dS_2}=\frac{1}{2\pi l}.
\end{equation}
A more familiar coordinate for dS$_2$ can be obtained by $\tau\rightarrow 2
\tau l/\lambda,x\rightarrow \lambda(x/l+1)/2$ to get
\begin{equation}
ds^2=-\left(-\frac{x^2}{l^2}+1\right)d\tau^2 +\left(-\frac{x^2}{l^2}%
+1\right)^{-1}dx^2 + k^2 d\tilde{\phi}^2.
\end{equation}

\subsection{Local Thermodynamic Stability}

Now, let us analyze the local thermodynamical stability of dS$_{3}$ black
holes and show that, as with their other features, it parallels the behavior of higher-dimensional
black holes.

Similar to the analysis of \cite{Anninos:2010gh,Birmingham:2010mj} we can
study the stability of the black holes in the canonical ensemble and grand
canonical ensemble.

In the canonical ensemble, temperature and angular momentum are fixed while
the Helmoltz free energy is given by
\begin{equation}
F=E-TS,
\end{equation}%
where $E\equiv E_{BH},T\equiv T_{+}$ and $S\equiv S_{BH}$.
The heat capacity is
\begin{equation}
C_{\mathcal{J}}=\left( \frac{\partial E}{\partial T}\right) _{\mathcal{J}},
\end{equation}%
where we are keeping $\mathcal{J}$ fixed. For the non-rotating black hole,
we have that $E(T)=\pi ^{2}l^{2}T^{2}/G$ implying that $C_{\mathcal{J}%
=0}=2\pi ^{2}l^{2}T/G>0$. For the rotating black holes, instead,
\begin{equation}
E=\frac{\pi ^{2}l^2T^{2}}{2G}\times \frac{\eta (E)+1}{\eta (E)^{2}},\quad
l\eta (E)=\frac{\sqrt{E^{2}l^{2}+\mathcal{J}^{2}}}{E},
\end{equation}%
so that
\begin{equation}
C_{\mathcal{J}}=\left( \frac{2E\eta ^{2}}{T}\right) \times \frac{1}{2-\eta }~,
\end{equation}%
which implies that $C_{\mathcal{J}}>0$ whenever $\eta <2$, i.e. $|a|=|%
\mathcal{J}|/E<\sqrt{3}l$ since the factor in the bracket is non-negative.
Note that the heat capacity blows up at $\eta =2$. Thus, we conclude that in
the canonical ensemble, the black holes are stable for $|\mathcal{J}|<\sqrt{3%
}El$.

On the other hand, in grand canonical ensemble, the system is studied at a
fixed temperature and angular velocity, and the thermodynamics potential is
given by the Gibbs free energy
\begin{equation}
G=E- T S - \Omega \mathcal{J}.
\end{equation}
Local thermodynamic stability is analyzed by studying the Hessian of
derivative of the entropy with respect to the charges $H_{ij} \equiv \frac{%
\partial^2 S}{\partial y^i \partial y^j}$ where $y^i=(E, \mathcal{J})$. In
particular, given the entropy
\begin{equation}
S(E,\mathcal{J})= \pi \sqrt{\frac{2}{ G l}} \sqrt{ E l+\sqrt{E^2l^2+\mathcal{%
J}^2} }
\end{equation}
we have a $2\times 2$ matrix in $(E,\mathcal{J})$
\begin{equation}
H_{ij}= \left(
\begin{array}{cc}
H_{11} & H_{12} \\
H_{12} & H_{22}%
\end{array}
\right).
\end{equation}
The regions that are locally stable must satisfy
\begin{equation}
H_{11} <0,\quad H_{22}<0,\quad \det{H}>0,
\end{equation}
i.e. the entropy is at a maximum. In fact,
\begin{equation}
\det{H}=-\frac{l \pi^2 }{4 G \left(E^2 l^2+\mathcal{J}^2\right)^{3/2}}<0.
\end{equation}
Thus, similar to the case in Kerr-dS$_4$ in Einstein gravity analyzed in
\cite{Anninos:2010gh}, \textit{all} black holes are (locally) thermally
unstable in the grand canonical ensemble where we allow exchange of angular
momentum.

\subsection{Microstates counting}

Equations (\ref{rmkabl1})-(\ref{rmkabl2}) express very interesting
properties of black holes in BHT theory which invites us to investigate
dS/CFT numerology in this setup. To do this, let us be reminded of the fact
that the existence of black hole solutions (\ref{uds}) demands $m^{2}
l^{2}=-1/2$. In particular, it implies that the central charge is
\begin{equation}  \label{eq:centralcharge}
c^\pm =-\frac{3l}{G}.
\end{equation}

The entropy (\ref{astec}) associated to configuration can readily be
computed and written in a suggestive form as%
\begin{equation}
S_{++}=\pi\sqrt{-\frac{l}{G}\left( l\mathcal{M}+i\mathcal{J}\right) }+\pi%
\sqrt{-\frac{l}{G}\left( l\mathcal{M}-i\mathcal{J}\right) },
\end{equation}%
which, given that $L_{0}^{\pm }=-(l{\mathcal{M}}\mp i{\mathcal{J}})/2$, can
be written as follows%
\begin{equation}
S_{++}=2\pi \sqrt{\frac{|c^{+}|L_{0}^{+}}{6}}+2\pi \sqrt{\frac{|c^{-}|{L}%
_{0}^{-}}{6}},  \label{Cardy}
\end{equation}%
with $c^{\pm }$ given in Eq.~(\ref{eq:centralcharge}). This manifestly shows
that Cardy formula (\ref{Cardy}) exactly reproduces horizon entropy (\ref%
{astec}).

The absolute value appearing in the Cardy formula is of course non-standard.
Let us try to justify it as follows: Imagine a CFT$_{2}$ with real and
positive central charge $c^{\pm }$ but with Hamiltonian $\mathcal{M}$
bounded from above instead of from below (this requires sending $\ell
_{n}\rightarrow -\ell _{-n}$ with respect to the convention we currently
have). Imagine also that for some reason we want to call the state with
maximum energy `the vacuum'. Let us, for simplicity, consider the canonical
partition function without angular potential. At finite (positive) inverse
temperature $\beta $, the usual partition function Tr$e^{-\beta \mathcal{M}}$
does not exist. However, the quantity $Z=\mbox{Tr}e^{+\beta \mathcal{M}}$
can be defined, and it allows to compute, for instance, the energy of the
system. With $\mathcal{M}_{BH}\equiv -\mathcal{M}$, the partition function
is back at its usual form $Z=\mbox{Tr}e^{-\beta \mathcal{M}_{BH}}$, and with
$\mathcal{M}_{BH}=-\partial _{\beta }\log Z$, $S=(1-\beta \partial _{\beta
})Z$ we have $d\mathcal{M}_{BH}=TdS$. In terms of the original Hamiltonian,
we thus have $d\mathcal{M}=-TdS$. Of course, this is just saying that when
the energy is bounded from above, one can flip the sign of the temperature
to make it bounded from below. Furthermore, if $Z$ is the partition function
of a CFT$_{2}$ with positive central charge and $L_{0}=\bar{L}_{0}=l\mathcal{%
M}/2$, the usual derivation of the Cardy formula -- see e.g. Appendix of
\cite{1210.0539} -- will go through if one uses $L_{0}^{BH}=\bar{L}%
_{0}^{BH}=-l\mathcal{M}/2$, giving a Cardy entropy $S\sim \sqrt{cL_{0}^{BH}}%
/6+\sqrt{c\bar{L}_{0}^{BH}/6}=\sqrt{c(-L_{0})/6}+\sqrt{c(-\bar{L}_{0})/6}$.
Therefore, in that case, it is natural to have a minus sign in the entropy
formula.

Following \cite{Bousso:2001mw}, one can also try to identify the left and
right temperatures
\begin{equation}
T_{L,R}\equiv \frac{T_{++}}{1\pm il\Omega _{++}}.
\end{equation}
Using Eq.~(\ref{eq:entropyThermo}), we have
\begin{equation}
S_{++}=\frac{\pi ^{2}l}{3}|c^{\pm }|(T_{L}+T_{R}),  \label{eq:CardyCanonical}
\end{equation}%
which is just the canonical ensemble form of Cardy formula. We can see that,
in contrast to the AdS$_{3}$ hairy black hole case, where there exists a
extremal rotation limit for which one of these temperatures vanishes\cite%
{GOTT}, here we do not have such situation. It is quite puzzling and perhaps
interesting that the $T_{L,R}$ never vanishes except at the special point $%
\mathcal{M}=0$ where they \textit{both} vanish
(as discussed in Sec.~\ref{sec:ext}).

This consideration about the non-existence of a smooth extremal limit (with
non-zero entropy) brings our attention to a different question, namely the
one about the inner black hole mechanics recently discussed in \cite%
{CastroRodriguez}. The laws of inner black hole mechanics in particular
state that the product of entropies associated to all the horizons has to be
independent of the black hole mass. This follows from the level matching
condition in the underlying CFT$_{2}$ description of the geometry. Inner
black hole mechanics was shown to hold well in a large class of models, and
particularly in three dimensions \cite{Detournay1}. However, at first
glance, the product of entropies $S_{+}S_{-}$ for the geometry (\ref%
{HairRhat}) does depend on the mass. Nevertheless, understanding whether or not this actually constitutes a
counterexample to the statements of inner black hole mechanics would require
further analysis (see \cite{CITEcvetic, CITEale} for some subtlety in this
issue). On the one hand, the hairy (A)dS$_{3}$ black holes present quite
curious thermodynamics: the inner and outer first laws are not independent
 from each other. On the other hand, at least for the dS case, the solution does not possess a smooth extremal non-zero-entropy limit.

\section{Conclusions}
In this paper he have studied asymptotically de Sitter black holes in
three-dimensional massive gravity. These black holes exhibit several features that are reminiscent of higher-dimensional dS black
holes in GR. In particular, they possess a curvature
singularity at the origin and interesting thermodynamical properties both at
the black hole event horizon and at the cosmological horizon. Unlike the GR
black holes in dS$_{d>3}$ space, these dS$_{3}$ black holes are always in
thermal equilibrium with respect with the cosmological horizon of the space
that hosts them.

For the massive gravity theory in three-dimensional dS space, we studied the
asymptotic isometry group and showed that it is generated by two copies
of the local conformal algebra in two-dimensions. The algebra of
the charges associated to the asymptotic Killing vectors consists of two
copies of Virasoro algebra with negative central charge. We also defined the regularized version of the Brown-York stress-tensor and
use it to compute the conserved charges of the black holes.

Finally, we showed that a naive application of the Cardy formula in the dual
CFT$_{2}$ theory exactly reproduces the entropy of both the black hole and
the cosmological horizon. The fact that Cardy entropy formula matches the
entropy of black holes in the bulk of dS$_{3}$ is quite remarkable. In fact,
in addition to mass and angular momentum, these black holes are characterized by an extra
hair parameter. Then, the fact that all these parameters conspire in a way
the numerical matching holds is quite surprising. Nevertheless, while
highly-suggestive, this numerical coincidence is far from being a substantial evidence of dS/CFT. Still, having been able to
reproduce the entropy of both the black hole and the cosmological horizon in
this setting encourages us to study this type of correspondence further.

\begin{equation*}
\end{equation*}

\textbf{Acknowledgements:} The authors thank Dionysios Anninos, Tatsuo
Azeyanagi, Alejandra Castro, Geoffrey Comp\`{e}re, Tom Faulkner, Tom
Hartman, Diego Hofman, Andr\'{e}s Goya, Julio Oliva, Soonkeon Nam, Jong-Dae
Park, Shahin Sheikh-Jabbari, Andy Strominger, Sang-Heon Yi for useful
discussions and comments. S.~dB. is funded by the Fonds
National de la Recherche Scientifique (Belgium) and also acknowledges
Wallonie-Bruxelles International for financial support. S.~D. was supported
by the Fundamental Laws Initiative of the Center for the Fundamental Laws of
Nature, Harvard University. The work of G.~G. has been supported by ANPCyT,
CONICET, ICTP, NSF, and UBA. G.~N. was supported in part by DOE grant
DE-FG02-91ER40654 and the Fundamental Laws Initiative at Harvard.


\begin{thebibliography}{99}
\bibitem{Strominger} A. Strominger,
\textit{Black Hole Entropy from Near-Horizon Microstates},
JHEP \textbf{02} (1998) 009, [arXiv:hep-th/9712251].

\bibitem{Waves} E. Ay\'{o}n-Beato, G. Giribet, and M. Hassa\"{\i}ne,
\textit{Bending AdS Waves with New Massive Gravity},
JHEP \textbf{05} (2009) 029, [arXiv:0904.0668].

\bibitem{Lifshitz} E. Ay\'{o}n-Beato, A. Garbarz, G. Giribet, and M. Hassa%
\"{\i}ne, \textit{Lifshitz black hole in three dimensions},
Phys. Rev. \textbf{D80} (2009) 104029, [arXiv:0909.1347].

\bibitem{flatCG} A. Bagchi, S. Detournay, and D. Grumiller,
\textit{Flat-Space Chiral Gravity},
Phys. Rev. Lett. \textbf{109} (2012) 151301, [arXiv:1208.1658].

\bibitem{glenn} G. Barnich, A. Gomberoff, and H. A. Gonz\'{a}lez,
\textit{The Flat limit of three dimensional asymptotically anti-de Sitter spacetimes},
Phys. Rev. \textbf{D86} (2012) 024020 [arXiv:1204.3288].

\bibitem{glenn2} G. Barnich, \textit{Entropy of three-dimensional
asymptotically flat cosmological solutions}, [arXiv:1208.4371].

\bibitem{detournay} A. Bagchi, S. Detournay, R. Fareghbal, and J. Simon,
\textit{Holography of 3d Flat Cosmological Horizons}, [arXiv:1208.4372].

\bibitem{Warped} D. Anninos, W. Li, M. Padi, W. Song, and A. Strominger,
\textit{Warped AdS}$_{3}$\textit{\ Black Holes},
JHEP 03 (2009)\ 130, [arXiv:0807.3040].

\bibitem{1210.0539} S. Detournay, T. Hartman and D. Hofman,
\textit{Warped Conformal Field Theory},
Phys.Rev. \textbf{D86} (2012) 124018, [arXiv:1210.0539].

\bibitem{dSCFT} A. Strominger, \textit{The dS/CFT Correspondence},
JHEP \textbf{0110} (2001) 034, [arXiv:hep-th/0106113];
\textit{Inflation and the dS/CFT Correspondence}
JHEP \textbf{0111} (2001) 049, [arXiv:hep-th/0110087].

\bibitem{Hull:1998vg} C.~M.~Hull,
\textit{Timelike T duality, de Sitter space, large N gauge theories and topological field theory},
JHEP \textbf{9807} (1998) 021, [arXiv:hep-th/9806146].

\bibitem{Witten:2001kn} E.~Witten, \textit{Quantum gravity in de Sitter
space,} [arXiv:hep-th/0106109]. 

\bibitem{Klemm:2001ea} D.~Klemm,
\textit{Some aspects of the de Sitter / CFT correspondence},
Nucl.\ Phys.\ B \textbf{625}, 295 (2002) [hep-th/0106247].

\bibitem{Cacciatori:2001un} S.~Cacciatori and D.~Klemm,
\textit{The Asymptotic dynamics of de Sitter gravity in three-dimensions},
Class.\ Quant.\ Grav.\ \textbf{19}, 579 (2002) [hep-th/0110031].

\bibitem{Maldacena:2002vr} J.~M.~Maldacena,
\textit{Non-Gaussian features of primordial fluctuations in single field inflationary models},
JHEP \textbf{0305} (2003) 013, [arXiv:astro-ph/0210603].

\bibitem{Anninos:2011jp} D.~Anninos, G.~S.~Ng and A.~Strominger,
\textit{Future Boundary Conditions in De Sitter Space},
JHEP \textbf{1202} (2012) 032, [arXiv:1106.1175].

\bibitem{Ng:2012xp} G.~S.~Ng and A.~Strominger, \textit{State Operator
Correspondence in Higher-Spin dS/CFT,} [arXiv:arXiv:1204.1057].

\bibitem{Anninos:2011af} D.~Anninos, S.~A.~Hartnoll and D.~M.~Hofman,
\textit{Static Patch Solipsism: Conformal Symmetry of the de Sitter Worldline},
Class.\ Quant.\ Grav.\ \textbf{29} (2012) 075002, [arXiv:1109.4942].

\bibitem{Anninos:2012qw} D.~Anninos, 
\textit{De Sitter Musings},
Int.\ J.\ Mod.\ Phys.\ A \textbf{27} (2012) 1230013, [arXiv:1205.3855].

\bibitem{Marolf:2012kh} D.~Marolf, I.~A.~Morrison and M.~Srednicki, \textit{
Perturbative S-matrix for massive scalar fields in global de Sitter space,}
[arXiv:1209.6039]. 

\bibitem{Anninos:2011ui} D.~Anninos, T.~Hartman and A.~Strominger, \textit{%
Higher Spin Realization of the dS/CFT Correspondence,}
[arXiv:arXiv:1108.5735]. 

\bibitem{Anninos:2012ft} D.~Anninos, F.~Denef and D.~Harlow, \textit{The
Wave Function of Vasiliev's Universe - A Few Slices Thereof,}
[arXiv:1207.5517]. 

\bibitem{Anninos:2013rza} D.~Anninos, F.~Denef, G.~Konstantinidis and
E.~Shaghoulian, \textit{Higher Spin de Sitter Holography from Functional
Determinants,} [arXiv:1305.6321]. 

\bibitem{Banerjee:2013mca} S.~Banerjee, A.~Belin, S.~Hellerman,
A.~Lepage-Jutier, A.~Maloney, D.~Radicevic and S.~Shenker, \textit{Topology
of Future Infinity in dS/CFT,} [arXiv:1306.6629].

\bibitem{Ouyang:2011fs} P.~Ouyang, \textit{Toward Higher Spin dS3/CFT2,}
[arXiv:1111.0276]. 

\bibitem{Park:1998qk} M.~-I.~Park,
\textit{Statistical entropy of three-dimensional Kerr-de Sitter space},
Phys.\ Lett.\ B \textbf{440} (1998) 275, [arXiv:hep-th/9806119].

\bibitem{Bousso:2001mw} R.~Bousso, A.~Maloney and A.~Strominger,
\textit{Conformal vacua and entropy in de Sitter space},
Phys.\ Rev.\ D \textbf{65} (2002) 104039, [arXiv:hep-th/0112218].

\bibitem{NMG} E. Bergshoeff, O. Hohm, and P. Townsend,
\textit{Massive gravity in three dimensions},
Phys. Rev. Lett. \textbf{102} (2009)\ 201301, [arXiv:0901.1766].

\bibitem{NMG2} E. Bergshoeff, O. Hohm, and P. Townsend,
\textit{More on massive 3D gravity},
Phys. Rev. \textbf{D79} (2009) 124042, [arXiv:0905.1259].

\bibitem{OTT} J. Oliva, D. Tempo, and R. Troncoso,
\textit{Three-dimensional black holes, gravitational solitons, kinks and wormholes for BHT massive gravity},
JHEP \textbf{07} (2009) 011, [arXiv:0905.1545].

\bibitem{GOTT} G. Giribet, J. Oliva, D. Tempo, and R. Troncoso,
\textit{Microscopic entropy of the three-dimensional rotating black hole of BHT massive gravity},
Phys. Rev. \textbf{D80} (2009) 124046, [arXiv:0909.2564].

\bibitem{Gabadadze:2012xv} G.~Gabadadze, G.~Giribet and A.~Iglesias, \textit{%
New Massive Gravity on de Sitter Space and Black Holes at the Special Point,}
[arXiv:1212.6279]. 


\bibitem{Kwon:2011ey} Y.~Kwon, S.~Nam, J.~-D.~Park and S.~-H.~Yi,
\textit{Quasi Normal Modes for New Type Black Holes in New Massive Gravity},
Class.\ Quant.\ Grav.\ \textbf{28} (2011) 145006, [arXiv:1102.0138].

\bibitem{TMG} S. Deser, R. Jackiw, and S. Templeton,
\textit{Three-dimensional massive gauge theories},
Phys. Rev. Lett. \textbf{48} (1982) 975.

\bibitem{OTT2} J. Oliva, D. Tempo, and R. Troncoso, 
\textit{Static spherically symmetric solutions for conformal gravity in three dimensions},
Int. J. Mod. Phys. \textbf{A24} (2009) 1588, [arXiv:0905.1510].

\bibitem{DeNe} S. Deser and R. Nepomechie,
\textit{Gauge invariance versus masslessness in de Sitter space},
Annals Phys. \textbf{154} (1984) 396.

\bibitem{Hi} A. Higuchi,
\textit{Forbidden mass range for spin-2 field theory in de Sitter space-time},
Nucl. Phys. \textbf{B282} (1987) 397.

\bibitem{DeWa} S. Deser and A. Waldron,
\textit{Gauge invariances and phases of massive higher-spins in (A)dS},
Phys. Rev. Lett. \textbf{87} (2001) 031601, [arXiv:hep-th/0102166];
\textit{Partial masslessness of higher spins in (A)dS},
Nucl. Phys. \textbf{B607} (2001) 577, [arXiv:hep-th/0103198].

\bibitem{BlCv} M. Blagojevic and B. Cvetkovic,
\textit{Extra gauge symmetries in BHT gravity},
JHEP \textbf{1103} (2011) 139, [arXiv:1103.2388].

\bibitem{Marolf} W. Kelly and D. Marolf,
\textit{Phase Spaces for asymptotically de Sitter Cosmologies},
Class. Quant. Grav. \textbf{29} (2012) 205013, [arXiv:1202.5347].

\bibitem{BH} J. Brown and M. Henneaux,
\textit{Central Charges In The Canonical Realization Of Asymptotic Symmetries: An Example From Three-Dimensional Gravity}
Commun. Math. Phys. \textbf{104} (1986) 207.

\bibitem{1305.1277} C. Krishnan and S. Roy, \textit{Higher Spin Resolution
of a Toy Big Bang}, [arXiv:1305.1277].


\bibitem{Skenderis:2002wp} K.~Skenderis,
\textit{Lecture notes on holographic renormalization},
Class.\ Quant.\ Grav.\ \textbf{19} (2002) 5849, [arXiv:hep-th/0209067].

\bibitem{Cunliff} C. Cunliff, \textit{Non-Fefferman-Graham asymptotics and
holographic renormalization in New Massive Gravity} JHEP \textbf{1304}
(2013) 141, [arXiv:1301.1347].

\bibitem{otro} S. Nam, J-D. Park, and S-H. Yi, 
\textit{Mass and angular momentum of black holes in new massive gravity},
Phys. Rev. \textbf{D82} (2010) 124049, [arXiv:1009.1962].

\bibitem{Balasubramanian} V. Balasubramanian, J. de Boer, D. Minic,
\textit{Mass, Entropy and Holography in Asymptotically de Sitter Spaces},
Phys. Rev. \textbf{D65} (2002) 123508, [arXiv:hep-th/0110108].

\bibitem{BB} G. Barnich and F. Brandt,
\textit{Covariant theory of asymptotic symmetries, conservation laws and central charges},
Nucl. Phys. \textbf{B633} (2002) 3, [arXiv:hep-th/0111246].

\bibitem{Jafferis:2013qia} D.~L.~Jafferis, A.~Lupsasca, V.~Lysov, G.~S.~Ng
and A.~Strominger, \textit{Quasinormal Quantization in deSitter Spacetime,}
[arXiv:1305.5523]. 

\bibitem{Anninos:2010gh} D.~Anninos and T.~Anous,
\textit{A de Sitter Hoedown},
JHEP \textbf{1008} (2010) 131, [arXiv:1002.1717].

\bibitem{dSCFT2} M. Spradlin, A. Strominger, and A. Volovich, \textit{Les
Houches Lectures on De Sitter Space}, [arXiv:hep-th/0110007].


\bibitem{Balasubramanian:1999re} V.~Balasubramanian and P.~Kraus,
\textit{A Stress tensor for Anti-de Sitter gravity},
Commun.\ Math.\ Phys.\ \textbf{208} (1999) 413, [arXiv:hep-th/9902121].

\bibitem{de Haro:2000xn} S.~de Haro, S.~N.~Solodukhin and K.~Skenderis,
\textit{Holographic reconstruction of space-time and renormalization in the AdS / CFT correspondence},
Commun.\ Math.\ Phys.\ \textbf{217} (2001) 595, [arXiv:hep-th/0002230].

\bibitem{HohmTonni} O. Hohm and E. Tonni, 
\textit{A boundary stress tensor for higher-derivative gravity in AdS and Lifshitz backgrounds},
JHEP \textbf{04} (2010) 093, [arXiv:1001.3598].

\bibitem{GiribetLeston} G. Giribet and M. Leston, 
\textit{Boundary stress tensor and counterterms for weakened AdS}$_{\mathit{3}}$\textit{\ asymptotic
in New Massive Gravity},
JHEP \textbf{09} (2010) 070, [arXiv:1006.3349].

\bibitem{Dio} D. Anninos and T. Hartman,
\textit{Holography at an Extremal De Sitter Horizon},
JHEP \textbf{03} (2010) 096, [arXiv:0910.4587].

\bibitem{Birmingham:2010mj} D.~Birmingham and S.~Mokhtari,
\textit{Thermodynamic Stability of Warped $AdS_3$ Black Holes},
Phys.\ Lett.\ B \textbf{697} (2011) 80, [arXiv:1011.6654].

\bibitem{CastroRodriguez} A. Castro and M.J. Rodr\'{\i}guez, \textit{%
Universal properties and the first law of black hole inner mechanics},
[arXiv:1204.1284].

\bibitem{Detournay1} S. Detournay, \textit{Inner Mechanics of 3d Black Holes}%
, [arXiv:1204.6088].

\bibitem{CITEcvetic} M. Cvetic, H. Lu and C. Pope, \textit{Entropy-Product
Rules for Charged Rotating Black Holes}, [arXiv:1306.4522].

\bibitem{CITEale} A. Castro, N. Dehmami, G. Giribet and D. Kastor, \textit{%
On the Universality of Inner Black Hole Mechanics and Higher Curvature
Gravity}, [arXiv:1304.1696].

\end{thebibliography}
\end{document}